\documentstyle[times,prl,aps,epsfig,amssymb,balanced]{revtex}
\draft
\begin{document}
\title{Stable droplets and growth laws close to the
modulational instability of a domain wall}
\author{Dami\`a Gomila$^{1,2}$, Pere Colet$^1$, Gian-Luca Oppo$^2$ and
Maxi San Miguel$^1$}
\address{$^1$Instituto Mediterr\'aneo de Estudios Avanzados, IMEDEA$^*$
(CSIC-UIB),\\
Campus Universitat Illes Balears, E-07071 Palma de Mallorca,
Spain.\\
$^2$Department of Physics and Applied Physics, University of Strathclyde,\\
107 Rottenrow, Glasgow G4 ONG, Scotland, UK.}
\maketitle
\begin{abstract}
We consider the curvature driven dynamics of a domain wall
separating two equivalent states in systems displaying a modulational
instability of a flat front. We derive an amplitude equation for the dynamics of
the curvature close to the bifurcation point from growing to schrinking circular
droplets. We predict the 
existence of stable droplets with a radius $R$ that diverges at the bifurcation 
point, where a curvature driven growth law 
$R(t) \approx t^{1/4}$ is obtained. Our general analytical predictions, which
are valid for a wide variety of systems including models of nonlinear optical
cavities and reaction-diffusion systems, are illustrated in the parametrically 
driven complex Ginzburg-Landau equation. 
\end{abstract}
\pacs{PACS numbers: 47.54.+r, 47.20.Ky, 42.65.Pc, 42.65.Sf}
\begin{twocolumns}
Growth of spatial domains of different phases
in the transient regime of a system approaching thermodynamic
equilibrium has been studied long ago \cite{gunton83}. This phenomenon
is a prototype of nonlinear dynamics of an extended system
governed by the motion of domain walls. Power laws for the growth
of a characteristic size and self-similar evolution (dynamical
scaling) have been established. Clear physical mechanisms
explaining the different asymptotic growth laws have also been identified 
\cite{gunton83}. It is convenient to consider domain growth in systems with 
conserved and non-conserved order parameters separately. In the first case one 
talks of spinodal decomposition leading to the Lifshits-Slyozov $t^{1/3}$ power 
law. In the case of non-conserved order parameters, the dominant mechanism is 
curvature driven minimization of surface tension, leading to the Allen-Cahn (AC)
$t^{1/2}$ power law \cite{Allen-Cahn}.

The study of domain growth and domain wall motion in non-equilibrium 
systems that do not approach thermodynamic equilibrium 
is by far more complicated. There is presently only a partial 
understanding of a variety of possible situations 
\cite{Meron,Cross95}. In particular a number of 
results have been recently reported for domain growth and domain wall 
motion in the transverse plane of nonlinear optical systems
\cite{Tlidi,Gallego,Oppo,Peschel,Leberre2}.
However, these numerical results often fail to identify the 
physical mechanism responsible for the observed dynamics. Moreover, 
the power laws for domain
growth do not always correspond unambiguously to an asymptotic and
self-similar  dynamics. An exception is the $t^{1/2}$ law obtained
for two different optical systems for which evidence of dynamical
scaling has been given \cite{Gallego,Oppo}. Visual inspection of the 
time dependent configurations clearly indicate curvature driven domain 
wall dynamics, but no derivation of such a law has been given for 
these systems. Surface tension, whose minimization may be used to 
explain the AC law, is not an appropriate concept for these 
optical systems. 

In this letter we analyze the transition from a regime characterized by a 
$t^{1/2}$ growth law to one of labyrinthine pattern formation due to a
modulational instability 
of a flat domain wall connecting two equivalent homogeneous solutions. Such a
transition from coarsening to labyrinthine regimes has been observed 
experimentally in reaction-diffusion \cite{Swinney} and in optical 
\cite{Staliunas} systems, and numerically
in \cite{Gallego}, as well as in Swift-Hohenberg models \cite{japos}. At
difference with systems approaching thermodynamic equilibrium, the coefficient
of the $t^{1/2}$ power law can now change sign upon variations of a
control parameter $p$ at $p=p_c$. For $p<p_c$ ( $p>p_c$) a circular domain of radius $R$ of
one solution embedded in the other grows (shrinks as in the AC regime) 
as $R(t) =\sqrt{R(0)- \gamma t}$ 
with $\gamma < 0$  ($\gamma > 0$). Moreover, for $p<p_c$ a flat front is
modulationally unstable and a generic initial condition develops in a
labyrinthine pattern. Close to the modulational instability at $p_c$ we derive
an amplitude equation for the curvature of circular domains. At $p_c$ a
circular domain grows as $R(t) \sim t^{1/4}$. For $p$ larger but close to
$p_c$ an initially small (very large) domain grows (shrinks) until a stable
droplet (SD) is formed. The radius of the SD diverges at $p_c$. The SD should
not be confused with other types of soliton-like localized structures (LS)
appearing in the AC regime when the front oscillatory tail 
prevents the domain collapse \cite{Oppo,Leberre,Gallego}.
The SD are also different from the critical nucleus of
nucleation theory \cite{gunton83} where the droplet is unstable 
and inner and outer solutions are not equivalent. Note also that droplets found
in reaction-difussion systems \cite{Goldstein} involve two non equivalent 
states.

Our analytical results are illustrated by considering a 
prototypical model, the parametrically driven complex Ginzburg-Landau 
equation (PCGLE)\cite{CoulletPRL,Coullet}, which describes a oscillatory system 
parametrically forced at twice its natural frequency \cite{Swinney}:
\begin{eqnarray}
\label{gleq}
\partial_t A &=&(1+i\alpha)\nabla^2 A+(\mu +i\nu)A \\ \nonumber
&&-(1+i\beta)|A|^2A+p A^* \,,
\end{eqnarray}
where $\mu$ measures the distance from the oscillatory instability threshold,
$\nu$ is the detuning parameter and $p>0$ is the forcing amplitude. For $p
\sim \nu \sim \alpha$ large compared to other parameters, a finite 
wavelength instability forming hexagons takes place for $p<p_h$, while
for $p>p_h$ there are two equivalent stable homogeneous solutions (frequency 
locked solutions)\cite{Coullet}.

We consider a real $N$ component vector field $\vec{\Psi}(\vec{x})$
whose dynamical evolution in two spatial dimensions is 
\begin{equation}
\label{equation}
\partial_t \vec{\Psi} = D \nabla^2 \vec{\Psi} + \vec{W}(\vec{\Psi},p) \,,
\end{equation}
where the matrix $D$ describes the spatial coupling,
$\vec{W}$ is a local nonlinear function of the fields and $p$ a control
parameter. Eq.\ (\ref{equation}) is invariant under translations and under the 
change  $\vec{x} \rightarrow -\vec{x}$. In addition we assume that it has a 
discrete symmetry $\mathcal{Z}$ that allows for the 
existence of two, and only two, equivalent stable homogeneous solutions, and 
that, in a 1d system, they are connected by a stable Ising front 
$\vec{\Psi}_0(x,p)$. An Ising front satisfies 
$\vec{\Psi}_0(x_0-x)={\mathcal Z}\vec{\Psi}_0(x-x_0)$ where $x_0$ is the 
front location \cite{Michaelis}, thus the 1d front 
(and equivalently a flat front in 2d) is stationary, 
$D \nabla^2 \vec{\Psi}_0 + \vec{W}(\vec{\Psi}_0,p)
=0$. For the PCGLE $ \vec{\Psi}(\vec{x})=(Re[A(\vec{x})], Im[A(\vec{x})])$,
$D=((1,\alpha)^T,(-\alpha,1)^T))$, $\vec{\Psi}_0(x)$ describes a flat front 
connecting the two homogeneous solutions (we consider $p>p_h$) 
and ${\mathcal Z}=-I$ where $I$ is the identity matrix.

Let $\vec{X}(s,t)$ represent the instantaneous position vector of the
front in the $\vec{x}$ plane, where $s$ is the arclength. It is convenient
to define a coordinate system $(r,s)$ that moves with the front such that
$\vec{x}=\vec{X}(s,t)+r\hat{r}(s,t)$,
where $\hat{r}$ is a unit vector normal to the curve $\vec{X}$ and the
coordinate $r$ is the distance of the point $\vec{x}$ to the front \cite{Meron}.
In the moving frame Eq.\ (\ref{equation}) becomes
\begin{eqnarray}
\label{equationmovingframe}
 D \partial_r^2 \vec{\Psi} + &&\left(vI+{\kappa \over 1+r\kappa}D \right)
\partial_r \vec{\Psi}  +{\kappa^2 D  \partial_{\theta}^2 \vec{\Psi} \over 
(1+r\kappa)^2} \nonumber \\
&& + \vec{W}(\vec{\Psi},p) =  \partial_t \vec{\Psi} \, ,
\end{eqnarray}
where $v=\partial_t\vec{X} \cdot \hat{r}$ is the (normal) front velocity
and $\theta=\kappa s$ is the azimuthal angle. We analyze the dynamics of 
slightly curved fronts as a perturbation of the flat front 
$\vec{\Psi}(r,s,t)=\vec{\Psi}_0(r)+\vec{\Psi}_1(r,s,t)$.
We assume that i) $\kappa w\ll 1$, with $\kappa =\nabla \cdot \hat{r}$ the
curvature and $w$ the front width, 
ii) in the moving frame the front profile depends at most weakly on t
($|\partial_t \vec{\Psi}| \ll |\kappa D \partial_r\vec{\Psi}|$) and iii) 
$\kappa$ is a function which depends at most weakly on $s$, thus $|\kappa
\partial_{\theta}^2 \vec{\Psi}| \ll |\partial_r\vec{\Psi}|$. Linearizing 
around $\vec{\Psi}_0$ we have
\begin{equation}
\label{1st order}
M \vec{\Psi}_1=-(v I +\kappa D)\partial_r \vec{\Psi}_0 \,,
\end{equation}
where $M^i_j=D^i_j \partial_r^2 + \delta_{\Psi^j} W^i 
\arrowvert_{\vec{\Psi}_0,p}$.
Due to the translational invariance of (\ref{equation}) $M$ is singular,
$M\vec{e}_0 = 0$ where $\vec{e}_0 \equiv \partial_r\vec{\Psi}_0$ is the
Goldstone mode. The solvability condition applied to (\ref{1st order}) 
leads to $v=-\gamma(p)\kappa$, where 
\begin{equation}
\label{gamma}
\gamma(p) \equiv {1\over \Gamma}\int_{-\infty}^\infty \vec{a}_0 \cdot D 
\vec{e}_0 dr \,, 
\end{equation}
$\Gamma \equiv \int_{-\infty}^\infty \vec{a}_0 \cdot \vec{e}_0 dr$ 
\cite{footnote1} and $\vec{a}_0$ is the null mode of $M^\dagger$. For a circular
domain $\kappa = 1/R$ and
\begin{equation}
\label{rdot}
v=\dot{R}=-\gamma(p)/R \,.
\end{equation}
From Eq.\ (\ref{1st order}) one obtains that the perturbation of the front,   
$\vec{\Psi}_1(r,t)=\kappa (t) \vec{\varphi}_1(r)$, is independent of $s$ while
the dependence on $t$  comes only through the curvature. $\vec{\varphi}_1$
satisfies
\begin{equation}
M \vec{\varphi}_1 = -(-\gamma I+D)\vec{e}_0.
\label{phisub1}
\end{equation}

For systems such that the diffusion matrix is proportional to the identity, 
$D=dI$, $\gamma$ takes the constant 
value $d$ independently of the profile of the front and any system 
parameter. In this case, the rhs of (\ref{phisub1}) vanishes and  
$\vec{\varphi}_1$ must be either zero or proportional to the Goldstone mode 
$\vec{e}_0$. Physically this means that the fronts translate without changing 
their radial profile. 
The front velocity is proportional to the curvature with opposite
sign $v=-d\kappa$, which is the well known AC law 
\cite{gunton83,Allen-Cahn}. This law implies a coarsening
regime with a $t^{1/2}$ growth law and shrinking of circular domains.

In general, $D=dI+C$ with a non zero matrix $C$. $C$ leads to a 
contribution to $\gamma$ that depends on the profile of the front $\vec{\Psi}_0$
and therefore, on the system parameters. From Eq.\ (\ref{phisub1}) we have
that $\vec{\varphi}_1$ is no longer proportional to the 
Goldstone mode. This means that the transverse profile of the front is now 
deformed. Since $\vec{\Psi}_1(r,t)=\kappa (t) \vec{\varphi}_1(r)$, 
the amount of deformation is proportional to the curvature and $C$. 
The $d$ contribution is generally positive and, for wide parameter 
regions, $\gamma$ is also positive. Here flat walls are stable and the 
AC law still applies \cite{Gallego,Oppo}.
Circular domains shrink but the presence of oscillatory tails in the front may
prevent the droplet to disappear forming a LS as found in 
nonlinear optical cavities \cite{Oppo,Leberre,Gallego}.

The crucial point is that for some parameter values the contribution 
to $\gamma$ due to $C$ may be negative and larger than $d$. 
$\gamma$ changes sign and a bifurcation occurs.
This is particularly relevant in nonlinear optical systems where the 
spatial coupling is diffractive and $d=0$.
For $\gamma<0$ the velocity has the same sign of the curvature leading to 
the growth of any perturbation of the flat wall. The condition for vanishing 
$\gamma$, $\int_{-\infty}^\infty \vec{a}_0 \cdot D \vec{e}_0 dr=0$, is in 
fact the criterion for the modulational instability of a flat front. 
Modulational instabilities in fronts connecting two equivalent homogeneous 
states have been observed in different systems
\cite{Gallego,Peschel}. Starting from a random initial condition, the
system develops labyrinthine patterns \cite{Gallego,Swinney}. Also, a 
circular domain of radius $R$ grows like (\ref{rdot}) until its boundary 
breaks up because of the modulational instability leading to the formation of a
labyrinthine pattern.

In Fig.\ref{gammap} we show the value of $\gamma$ for the PCGLE 
calculated using definition (\ref{gamma}), with 
the profile of the 1d front calculated numerically by solving the stationary
form $\partial_tA=0$ in 1d of (\ref{gleq}). This calculation identifies
$p_c$ for which $\gamma =0$.

We are now ready to obtain an amplitude equation for the curvature in the
vicinity of $p_c$. We start considering the case of a circular domain wall
(for which $v=-\dot{\kappa}/\kappa^2$). Close to $p_c$ ($\gamma \approx 0$) we
perform a multiple scale analysis in $\epsilon$ of Eq.\ 
(\ref{equationmovingframe})
with $p=p_c + \epsilon p_1$, $\vec{\Psi}=\vec{\Psi}_0+\epsilon^{1/2}\vec{\Psi}_1
+\epsilon\vec{\Psi}_2 +\epsilon^{3/2}\vec{\Psi}_3$, $\kappa=\epsilon^{1/2}
\kappa_1$ and $\partial_t=\epsilon^{2}\partial_T$.
At order $\epsilon^{1/2}$ we obtain
$M\vec{\Psi}_1=  -\kappa_1 D \vec{e}_0$. The solvability
condition implies $\int_{-\infty}^\infty \vec{a}_0 \cdot D \vec{e}_0
dr=0$, which is automatically satisfied at $p=p_c$. One finds 
$\vec{\Psi}_1=\kappa_1 \vec{\varphi}_1$ where
$\vec{\varphi}_1$ satisfies $M\vec{\varphi}_1=-D \vec{e}_0$.
Due to the symmetry of the front the solvability condition at order $\epsilon$
is always fulfilled and we get 
$\vec{\Psi}_2=p_1 \vec{\varphi}_2 + \kappa_1^2
\vec{\varphi}_3$ with $M \vec{\varphi}_2=- \partial_{p}\vec{W}|_0$ and 
$M^i_j \varphi_3^j=-D^i_j(\partial_r \varphi_1^j-r e_0^j)-{1\over 2} 
\delta_{\Psi^j\Psi^k}W^i|_0\varphi_1^j\varphi_1^k$
where $|_0$ means evaluated at $\vec{\Psi}_0$ and $p_c$.
At order $\epsilon^{3/2}$, the amplitude equations for the 
curvature $\kappa_1$ and the radius of the circular droplet
\begin{eqnarray}
\label{amplitudeeq}
{\partial_T \kappa_1 \over \kappa_1^2}&&=c_1p_1 \kappa_1+c_3 \kappa_1^3 \\
\partial_t R=-&&c_1(p-p_c){1\over R} -c_3 {1\over R^3} 
\label{eqrapidvariables}
\end{eqnarray}
where 
\begin{eqnarray}
&c_1&={1 \over \Gamma} \int_{-\infty}^\infty a_{0i}
(D^i_j\partial_r \varphi_2^j 
+ \delta_{\Psi^j} \partial_pW^i|_0 \varphi_1^j+ \nonumber \\ 
&& \delta_{\Psi^j\Psi^k} W^i|_0\varphi_1^j\varphi_2^k)dr \\
&c_3&={1 \over \Gamma} \int_{-\infty}^\infty a_{0i} 
[D^i_j(\partial_r \varphi_3^j-r\partial_r\varphi_1^j+r^2 e_0^j)+ \nonumber \\ 
&&\delta_{\Psi^j\Psi^k}W^i|_0\varphi_1^j\varphi_3^k+{1\over 6}
\delta_{\Psi^j\Psi^k\Psi^l} W^i|_0\varphi_1^j\varphi_1^k\varphi_1^l] dr 
\label{c3}
\end{eqnarray}
are obtained from the solvability condition.
The coefficient $c_1$ is positive since $\gamma=c_1 (p-p_c)$, and we are
considering $\gamma>0$ for $p>p_c$.
If $c_3$ is negative (supercritical bifurcation) our analysis predicts just 
above the modulational instability the existence of stable stationary
circular domains (SD) with a very large radius 
\begin{equation}
\label{R0}
R_0={1\over \sqrt{p -p_c}}\sqrt{-c_3\over c_1}.
\end{equation}
In Fig. \ref{cdw3d} (left) we show the form of the SD for the PCGLE.
At the center the field takes the value of one of the homogeneous solutions, so
the wall of this structure is a heteroclinic orbit between the two homogeneous
states. The radius of the SD can be extremely large 
diverging at $p=p_c$. Fig. \ref{cdw3d} (right) displays the radius of the 
SD and that of the LS calculated 
solving numerically $(1+i\alpha)(\partial_r^2+{1\over r}\partial_r) A
+(\mu +i\nu)A 
-(1+i\beta)|A|^2A+p A^*=0$. The inset in Fig. \ref{cdw3d} (right) shows the 
 linear dependence of $1/R_0^2$ with $p$ as predicted by (\ref{R0}). 
In spite of the fact that there is a smooth 
transition from LS to SD, they are intrinsically different. 
The interaction of the oscillatory tails which is responsible for the existence
of LS does not play any role in the SD. The 
stabilization mechanism comes from the counterbalance between
the $R^{-3}$ contribution to the front velocity and the shrinking due to 
the $R^{-1}$ contribution.

At the bifurcation point $p=p_c$, Eq.\ (\ref{eqrapidvariables})
becomes $\partial_t R = -c_3 / R^3$ and any circular domain of one
solution embedded in the other grows as $R(t) \sim
t^{1/4}$. In Fig. \ref{R4t} we show the time evolution of the radius
of a circular domain for the PCGLE at $p=p_c$. It fits nicely
the theoretical dependence we predict. Close to $p_c$, 
in the regime of existence of the SD, there is no asymptotic power law of 
domain growth since at very long times the SD is formed stopping the
growth process. During the transient, an initially small (very large) circular 
domain will grow (shrink) following (\ref{eqrapidvariables}).

So far we have considered the dynamics of domains with radial
symmetry. When the system evolves from random initial conditions
other dynamical mechanisms come into play.  
Close to the bifurcation point the main non-radially symmetric contribution to
the velocity comes from the variation of the curvature along the front.
The derivation of the amplitude equation for the curvature leads to an 
additional term in Eq.\ (\ref{amplitudeeq}) 
$c_2 \kappa_1^2 \partial_{\theta}^2 \kappa_1$
where $c_2= {1\over \Gamma}\int_{-\infty}^\infty \vec{a_0}\cdot D 
\vec{\varphi}_1 dr$. The front velocity becomes
$v=-c_1(p-p_c)\kappa -c_2 \kappa^2 
\partial_{\theta}^2 \kappa -c_3 \kappa^3$.
Consistently with our approximations, $\partial_{\theta}^2 \kappa$ will
change at most at order $\kappa^0$, so the non-radial contribution is 
at least of order $\kappa^2$.
Close to the bifurcation point the term proportional to  $\kappa$ can be 
neglected, and the term given by
$-c_2\kappa^2 \partial_{\theta}^2 \kappa$ might be dominant compared with 
the one proportional to $\kappa^3$. In this case the front velocity 
is proportional to $\kappa^2$. However, for any closed boundary 
$\partial_{\theta}^2 \kappa$ is positive in some parts of the wall and negative
in others, so this term does not lead to an asymptotic growth law. For $c_2 <0$,
which is the case for the PCGLE, this term tends to reduce the curvature
differences, so at $p_c$ an arbitrarily shaped domain first becomes circular
until the contribution of $\partial_{\theta}^2 \kappa$ vanishes and then the
circular domain grows as $R(t)\sim t^{1/4}$ due to the $c_3$ term 
(see Fig.\ \ref{mallorca}). 

In summary, we have analyzed a generic situation of domain wall
motion driven by curvature effects in which the proportionality
coefficient $\gamma$ between wall velocity and curvature changes
sign at a bifurcation point. In optical systems, this change is a
consequence of the diffractive coupling between real and imaginary
parts of the complex field amplitude. The amplitude equation for
the curvature in the vicinity of the bifurcation point predicts
the existence of stable nonlinear solutions which
are droplets of one phase embedded in a background of the second
equivalent phase. Nonlinear dynamics of the curvature leads to
growth laws different from the AC $t^{1/2}$-growth law.
The existence of a large characteristic length given by the radius
of the SD destroys the possibility of self-similar evolution.

The authors acknowledge financial support from the EC TMR Network 
QSTRUCT (FMRXCT960077). GLO acknowledges SGI and EPSRC 
(grants M19727 and M31880) for financial support. DG, PC and MSM 
acknowledge financial support from MCyT (Spain, projects PB97-0141-C02-02 and 
BFM2000-1108) and helpful discussions with E. Hern\'andez-Garc\'\i a.

\begin {figure}
\centerline{\psfig{figure=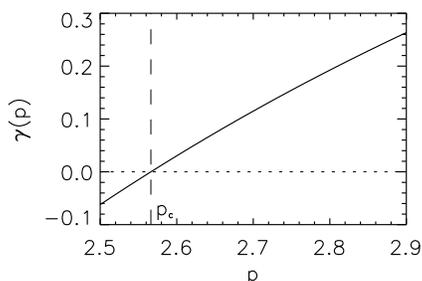}}
\caption[]{Value of the $\gamma$ as a function of the forcing amplitude $p$ for
the PCGLE calculated from (\ref{gamma}). We take here $\alpha=2$, $\beta=0$, $\nu=2$ and $\mu=0$. For these
values of the parameters $p_h=2.09$ and $p_c=2.56629$.}
\label{gammap}
\end{figure}
\begin {figure}
\centerline{\psfig{figure=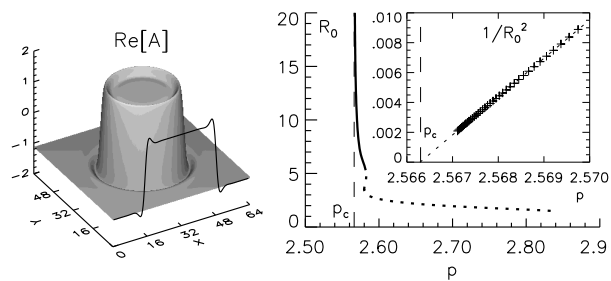}}
\caption[]{Left: Spatial dependence and transverse section of a SD for the 
PCGLE. Right: Radius of SD (solid line) and LS (dotted line) as 
a function of the forcing for the PCGLE. The inset shows
the linear dependence of $1/R_0^2$ with $p$ close to the bifurcation point
as predicted by (\ref{R0}).}
\label{cdw3d}
\end{figure}
\begin {figure}
\centerline{\psfig{figure=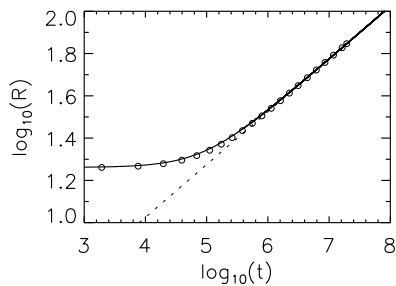}}
\caption[]{Growth of a circular domain as function of the time at $p=p_c$. The
symbols correspond to the numerical integration of Eq.\ (\ref{gleq}), the solid 
line is the theoretical prediction (\ref{eqrapidvariables}) with
$c_3=0.3129$ calculated from (\ref{c3}) and the dotted line has a slope 1/4 
showing the asymptotic behavior.}
\label{R4t}
\end{figure}
\begin {figure}
\centerline{\psfig{figure=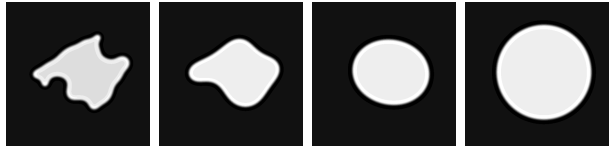}}
\caption[]{Evolution of an arbitrarily shaped domain at $p=p_c$. From left to
right: t=0, t=2400, t=80000 and t=1650000.}
\label{mallorca}
\end{figure}

\end{twocolumns}

\begin{thebibliography}{99}
\bibitem{footnote1} $\Gamma$ vanishes at
a Ising-Bloch transition \cite{Michaelis}. Here we only consider parameter 
regions far away from any Ising-Bloch transition for which $\Gamma$  is never 
zero.
\end{thebibliography}
\end{document}